\documentclass[11pt,a4paper]{article}
\pdfoutput=1
\usepackage{jheppub}
\usepackage{latexsym,amsfonts,amsmath,amssymb} 
\usepackage{bbm}
\usepackage{graphicx}
\usepackage{color}
\usepackage{caption}
\usepackage{subcaption}
\usepackage[normalem]{ulem} 
\usepackage{comment}
\usepackage{url}

\allowdisplaybreaks 
\def\Slash#1{\rlap{\hbox{$\mskip 3 mu /$}}#1}      

\renewcommand{\d}{\delta}
\renewcommand{\a}{\alpha}

\newcommand{\qeq}{\delta_\text{eq}}

\newcommand{\CL}{\mathcal{L}}
\newcommand{\CN}{\mathcal{N}}
\newcommand{\CV}{\mathcal{V}}

\newcommand{\bea}{\begin{eqnarray}}
\newcommand{\eea}{\end{eqnarray}}
\newcommand{\be}{\begin{equation}}
\newcommand{\ee}{\end{equation}}
\renewcommand{\=}{\;  = \;}

\title{BRST quantization and equivariant cohomology: 
localization with asymptotic boundaries}

\author{Bernard de Wit$^{a,b}$,} 
\author{Sameer Murthy$^c$,}
\author{and Valentin Reys$^{d,e}$}

\affiliation{$^a$ Nikhef Theory Group, Science Park 105, 1098 XG
  Amsterdam, The Netherlands}
\affiliation{$^b$ Institute for Theoretical Physics, Utrecht University, \\
  Leuvenlaan 4, 3584 CE Utrecht, The Netherlands}
\affiliation{$^c$ Department of Mathematics, King's College London,\\
  The Strand, London WC2R 2LS, U.K}
\affiliation{$^d$ Dipartimento di Fisica, Universit\'a di Milano-Bicocca,\\
  Piazza della Scienza 3, I-20126 Milano, Italy}
\affiliation{$^e$ INFN, Sezione di Milano-Bicocca,\\
  Piazza della Scienza 3, I-20126 Milano, Italy.}

\emailAdd{b.dewit@uu.nl}
\emailAdd{sameer.murthy@kcl.ac.uk}
\emailAdd{valentin.reys@unimib.it}

\abstract{We develop BRST quantization of gauge theories with a soft
  gauge algebra on spaces with asymptotic boundaries. The asymptotic
  boundary conditions are imposed on background fields, while quantum
  fluctuations about these fields are described in terms of quantum
  fields that vanish at the boundary.  This leads us to construct a
  suitable background field formalism that is generally applicable to
  soft gauge algebras, and therefore to supergravity.  We define a
  nilpotent BRST charge that acts on both the background and the
  quantum fields, as well as on the background and quantum
  ghosts.\\
  When the background is restricted to be invariant under a residual
  isometry group, the background ghosts must be restricted accordingly
  and play the role of the parameters of the background
  isometries. Requiring in addition that the background ghosts will
  be BRST invariant as well then converts the BRST algebra into an
  equivariant one. The background fields and ghosts are then invariant
  under the equivariant transformations while the quantum fields and
  ghosts transform under both the equivariant and the background
  transformations.  We demonstrate how this formalism is suitable for
  carrying out localization calculations in a large class of theories,
  including supergravity defined on asymptotic backgrounds that admit
  supersymmetry.  }

\begin{document}

\maketitle

\section{Introduction}
\label{sec:introduction}

The standard quantization of gauge theories, especially in the context
of perturbation theory, is carried out by imposing suitable gauge
conditions that require the introduction of so-called ghost fields
\cite{Feynman:1963ax,DeWitt:1967ub,Faddeev:1967fc}. The theory is then
no longer invariant under local gauge transformations, but under a
rigid fermionic nilpotent variation $\delta_\mathrm{brst}$ known as
BRST symmetry \cite{Becchi:1974md,Tyutin:1975qk}. When the generators
of the gauge group close under commutation, the quantum action
involves terms that are bilinears of ghost and anti-ghost fields. 
The BRST variations of the original fields can be directly expressed
in terms of the original gauge transformations with their parameters
replaced by the ghost fields.  The partition function for BRST
invariant operators as well as the S-matrix are then independent of
the gauge condition.\footnote{ 
  When the gauge algebra closes only modulo the equations of motion,
  then additional terms will be required of higher order in the ghost
  and anti-ghost fields in both the action and the BRST
  transformations. In that case one is dealing with an {\it open} BRST
  algebra
  \cite{Fradkin:1977xi,deWit:1978hyh,Batalin:1981jr,Batalin:1983wj}. For
  a review, see \cite{Gomis:1994he}.  } 

The formal structure of BRST transformations can in certain cases also
be used in the study of topological theories, where one has a
nilpotent fermionic operator~$\d$, often arising as a twisted
supercharge of some supersymmetric
theory~\cite{Witten:1988ze,Witten:1988xj}. Here the ghosts will
usually not originate from quantizing the theory, but they are
provided by the matter fermions of the original theory. The functional
integral then localizes to the $\delta$-cohomology.  More generally,
one can consider a fermionic symmetry operator~$\delta_\mathrm{eq}$
with algebra $\delta_\mathrm{eq}{\!}^2 = \delta_{\mathring{\xi}}$,
where~$\delta_{\mathring{\xi}}$ is the generator of a compact bosonic
symmetry. In this case one can apply the powerful mathematical
framework of {\it equivariant
  localization}~\cite{Duistermaat:1982vw,Berline:1982,Atiyah:1984px},
with the result that the functional integral will localize to
the~$\delta_\mathrm{eq}$-cohomology. This technique has been used to
great effect in the context of supersymmetric gauge
theories~\cite{Nekrasov:2002qd}, by choosing $\delta_{\mathring{\xi}}$
to be a combination of a compact isometry and internal symmetry
variations. These techniques can be extended to supersymmetric
theories on curved manifolds admitting non-trivial rigid
symmetries~\cite{Pestun:2007rz}.

These developments have led to a large number of applications, but
essentially all of them deal with \emph{rigid} supersymmetry (see the
review~\cite{Pestun:2016zxk} for a collection of recent results).  In
this paper we lay out a formalism for \emph{local} supersymmetry,
which can account for the fluctuations of (super-)gravitons in
the path integral.  We were motivated to study this problem in the
context of applying localization to determine the exact entropy of BPS
black holes \cite{Sen:2008vm} in supergravity---a program which has
already produced interesting
results~\cite{Dabholkar:2010uh,Dabholkar:2011ec}, but where the
underlying formalism needs to be put on a more rigorous footing. Hence
the focus in this paper will be on the complications that one
encounters when attempting to apply localization to theories with
fluctuating (super-)gravitons.

The power of the localization method is that it reduces an
infinite-dimensional functional integral to an integral over
$\qeq$-invariant field configurations. This is an enormous reduction
which, in lucky situations, could even lead to a finite-dimensional
integral.  Field configurations that are~$\qeq$-invariant are
necessarily $\delta_{\mathring{\xi}}$-invariant, and an appropriate
choice of the {\it background} bosonic
symmetries~$\delta_{\mathring{\xi}}$ constrains the
field configurations to fluctuate only along a restricted set of
directions in space-time as well as in field space.  In theories of
supergravity, however, the meaning of~$\qeq$
and~$\delta_{\mathring{\xi}}$ are not a priori clear, as both
supersymmetry as well as space-time translations are part of the gauge
algebra of supergravity.
One situation in which we can make sense of a rigid symmetry in a
gravitational theory is to consider a space with
a boundary and fix the behavior of all the fields near the
boundary.\footnote{
  In the context of AdS/CFT this is particularly natural, and, as is
  well-known, the space of boundary configurations of the bulk
  gravitational theory couples to the non-gravitational theory and
  thus inherits its rigid symmetries.} 
The functional integral is then performed over all the 
field fluctuations about a fixed background field configuration that
satisfies the boundary conditions.

A first natural step in this situation would be to recast the problem
in the background field formalism.  In trying to work out the details,
however, we run into a technical hurdle, namely that a general
understanding of the background field method is lacking for gauge
theories with \emph{soft algebras}, i.e.~theories in which the structure
`constants' of the gauge algebra are functions of fields (as is the
case for supergravity).  We solve this problem by constructing a
nilpotent BRST operator for soft gauge algebras in a situation where
the fields have been split into background and quantum pieces, and by
introducing two corresponding sets of ghosts.  The BRST operator then
acts on both the classical and the quantum fields, as well as on the
two sets of ghosts.  Subsequently we consider a
functional integral that only depends on the background fields (but not,
as it turns out, 
on the background ghosts), which is gauge independent provided the
background fields are invariant.  
As a next step we deform the BRST operator to an
equivariant symmetry~$\qeq$, by appropriately freezing the background
fields and ghosts, so as to obtain a rigid supersymmetry algebra of
the boundary, with an action on the full space of classical as well as
quantum fields. Our construction is very general in that it provides a
framework for equivariant localization for any gauge algebra
(including soft algebras) with some choice of a rigid subalgebra that
is picked by the boundary.

At a technical level, our problem involves setting up the action
of~$\qeq$ on the set of all fields in the gauge-fixed theory, and
computing the~$\qeq$-cohomology.  Different methods have been used in
the past to solve this cohomology problem, including BRST-based
methods~\cite{Baulieu:1988xs}. We refer the reader
to~\cite{Szabo:2000fs} for a comprehensive review. The main new points
that we discuss in this paper are functional integrals in theories
with soft gauge algebras, and the general mechanism of how background
symmetries act on quantum fields.  The application to localization in
supergravity can then be accomplished by specializing to a subalgebra
of the background isometries that contains a supersymmetry which
squares to a compact background isometry.
We then show that the functional integral localizes to the
space of~$\qeq$-invariant field configurations. This should, for
instance, enable one to carry out a first-principles calculation of the
exact quantum entropy of half-BPS black holes in~$\mathcal{N}=2$
supergravity, thus completing the analysis
of~\cite{Dabholkar:2010uh,Dabholkar:2011ec}. Although the present
paper is inspired by thinking about localization for BPS black holes
in supergravity, we should stress that we present a rather general
formalism that can equally well be used in a broader context.

The plan of the paper is as follows. In Section~\ref{sec:brst-general}
we present a brief review of BRST quantization for soft gauge algebras
and establish the notation.  For simplicity we restrict ourselves to
bosonic gauge invariances only, but at the end of the section we
indicate how to deal with the more general case.  Subsequently we
introduce the background field formulation in
Section~\ref{sec:back-field}. We define a functional integral that
only depends on the background fields and that is independent of the
gauge condition when the background fields are invariant. In
Section~\ref{sec:equiv-cohom} we discuss an equivariant cohomology
that arises upon a specific deformation where all the background
fields and ghosts are invariant and the quantum fields and ghosts
transform under $\delta_\mathrm{eq}$, which squares to a background
isometry $\delta_{\mathring{\xi}}$. Under certain conditions the
functional integral introduced in Section~\ref{sec:back-field} is also
invariant under this equivariant algebra. In the next
Section~\ref{sec:localization}, we demonstrate how this equivariant
algebra can be used for localization. Finally in
Section~\ref{sec:application} we present further details on how to
apply our method when determining BPS black hole entropy.

\section{BRST cohomology for soft algebras} 
\label{sec:brst-general}
\setcounter{equation}{0}

To introduce our notation we first define the BRST transformations in
the generic case of a gauge theory of bosonic gauge transformations
with a gauge algebra that closes off shell (i.e.~without the need of
imposing the field equations). Hence we have gauge transformations
expressed in terms of corresponding space-time dependent parameters
$\xi^\alpha(x)$. The infinitesimal gauge transformations of the fields
$\phi^i$ are written as follows,
\begin{equation}
  \label{eq:gauge-transf}
  \delta\phi^i \= R(\phi)^i{\!}_\alpha \, \xi^\alpha\,, 
\end{equation}
where $R(\phi)^i{\!}_\alpha$ may include derivatives acting on the
parameters $\xi^\alpha(x)$ and may depend non-linearly on the fields
$\phi^i$. They must satisfy the general closure relation
\begin{equation}
  \label{eq:closure2}
  \delta(\xi_1)\,\delta(\xi_2) -\delta(\xi_2)\,\delta(\xi_1) \=\delta(\xi_3)\,,
\end{equation}
with
$\xi_3{}^\alpha = f_{\beta\gamma}{}^\alpha \,\xi_1{}^\beta
\,\xi_2{\!}^\gamma$.
The structure `constants' $f_{\beta\gamma}{}^\alpha$ may depend on
$\phi^i$ and follow directly from the closure relation
\eqref{eq:closure2}. This leads to the following result, 
\begin{equation}
  \label{eq:closure}
   R^j{\!}_{[\alpha}\, \partial_j R^i{\!}_{\beta]} \= \tfrac12
   f_{\alpha\beta}{}^\gamma \, R^i{\!}_\gamma  \,.  
\end{equation}
Upon applying a third infinitesimal gauge transformation one 
derives the corresponding Jacobi identity,
\begin{equation}
  \label{eq:Jacobi}
    f_{[\alpha\beta}{}^\delta \,f_{\gamma]\delta}{}^\epsilon +
  R^j{\!}_{[\alpha}\,\partial_j f_{\beta\gamma]}{}^\epsilon  \=0\,. 
\end{equation}
Gauge algebras with field-dependent structure constants are often
called soft algebras. Supergravity theories are usually based on
a soft gauge algebra. The closure relation \eqref{eq:closure} and the
corresponding Jacobi identy \eqref{eq:Jacobi} will play an important
role throughout this paper.

The BRST transformations for the fields $\phi^i$ and the ghosts $c^\alpha$ then
take the following form,
\begin{align}
  \label{eq:BRST-tr}
  \delta_\mathrm{brst} \phi^i \=&\,R(\phi)^i{\!}_\alpha \,
  \Lambda\, c^\alpha
  \,, \nonumber\\
  \delta_\mathrm{brst} c^\alpha\=&\,  \tfrac12\,f_{\beta\gamma}{}^\alpha\,
  c^\beta\,\Lambda \,  c^\gamma  \,. 
\end{align}
Here we have introduced an auxiliary anti-commuting number $\Lambda$,
so that the fields and their variations have the same statistics. Its
presence also helps to keep track of the various minus signs that one
will encounter in the calculations.  It is straightforward to verify
that the above transformations are nilpotent when acting on $\phi^i$
or $c^\alpha$ by virtue of \eqref{eq:closure} and \eqref{eq:Jacobi},
\begin{equation}
  \label{eq:nilpotent}
  \delta_\mathrm{brst}{\!}^2 \,\phi^i \= 0\,,\qquad \delta_\mathrm{brst}{\!}^2\,
  c^\alpha \=0\,.
\end{equation}
To see this one applies two consecutive BRST transformations with
anti-commuting parameters $\Lambda_1$ and $\Lambda_2$.

The gauge-invariant classical Lagrangian
$\mathcal{L}^\mathrm{class} (\phi)$ is obviously BRST invariant,
because the BRST transformations on the fields $\phi^i$ take the form
of an infinitesimal gauge transformation with field-dependent
parameters.  We allow for an arbitrary Lagrangian of this type, which
may be formulated in space-times of various signatures.  In addition
we must include an extra BRST invariant term denoted by
$\mathcal{L}^\mathrm{g.f.}$ to fix the gauge, which will also provide
the ghost-dependent terms in the full quantum action. This requires
the introducion of anti-ghost fields $b_\alpha$ and Lagrange
multiplier fields $B_\alpha$, which will also transform under nilpotent
BRST transformations that we will define momentarily. The
invariance of the action $\mathcal{L}^\mathrm{g.f.}$ is then achieved
by writing it as a BRST transformation of the so-called {\it gauge
  fermion},
\begin{equation}
  \label{eq:Lagr-gf}
  \mathcal{L}^\mathrm{g.f.} \= \partial_\Lambda\,\delta_\mathrm{brst}
  \big[ b_\alpha\, F(\phi)^\alpha\big] \,. 
\end{equation}
When $\delta_\mathrm{brst}b_\alpha$ is proportional to $B_\alpha$,
then $B_\alpha$ will indeed act as a Lagrange multiplier for the gauge
choice $F(\phi)^\alpha=0$.  Note that we have extracted the auxiliary
anti-commuting number $\Lambda$ by a left derivative
$\partial_\Lambda$.

Choosing $\delta_\mathrm{brst} b_\alpha= \Lambda\,B_\alpha$ and
$\delta_\mathrm{brst} B_\alpha=0$, one ensures that the BRST
transformations on $b_\alpha$ and $B_\alpha$ are indeed
nilpotent. Subsequently one obtains the following expression 
for~$\mathcal{L}^\mathrm{g.f.}$,
\begin{equation}
  \label{eq:L-g-f}
  \mathcal{L}^\mathrm{g.f.} \= B_\alpha\,F(\phi)^\alpha -
  b_\alpha\,R(\phi)^j{\!}_\beta \,c^\beta\,\partial_j F(\phi)^\alpha\,,
\end{equation}
where we also assumed that the $\phi^i$ are commuting fields.  The
last term is precisely the Faddeev-Popov ghost Lagrangian
\cite{Faddeev:1967fc}.  Hence the BRST Lagrangian equals
\begin{equation}
  \label{eq:BRST-Lagrangian}
  \mathcal{L}_\mathrm{brst}(\phi^i,c^\alpha,b_\alpha,B_\alpha) \=
  \mathcal{L}^\mathrm{class}(\phi^i) +  \mathcal{L}^\mathrm{g.f.}(\phi^i,
  c^\alpha,b_\alpha,B_\alpha)  \,,
\end{equation} 
which is invariant under the combined BRST transformations 
\begin{equation}
  \label{eq:BRST-tr}
  \begin{array}{rcl}
    \delta_\mathrm{brst} \phi^i &\=& R(\phi)^i{\!}_\alpha \,
  \Lambda\, c^\alpha \,, \\
  \delta_\mathrm{brst} c^\alpha&\=&\,  \tfrac12\,f_{\beta\gamma}{}^\alpha\,
  c^\beta\,\Lambda \,  c^\gamma  \,, 
  \end{array} 
  \qquad
  \begin{array}{rcl} 
  \delta_\mathrm{brst} b_\alpha&\=&\,  \Lambda \, B_\alpha \,, \\
  \delta_\mathrm{brst} B_\alpha &\=&\, 0\,.
    \end{array} 
\end{equation}
The action corresponding to the Lagrangian \eqref{eq:BRST-Lagrangian}
can be used to define a corresponding path integral by integrating
over the various fields. Here it is important that the integral
measure is also invariant under the BRST transformations. The BRST
cohomolgy is based on the fact that the BRST transformations are
nilpotent on all the fields.

The quantities $F(\phi)^\alpha$ are known as the gauge-fixing terms
and ensure that the gauge invariance is removed. In principle this
implies that the number of degrees of freedom will change, because the
gauge fields will now acquire an additional degree of
freedom. However, at the same time we have included a Lagrange
multiplier field $B_\alpha$ of the same statistics as the
corresponding gauge field, as well as a ghost field $c^{\,\alpha}$ and
an anti-ghost field $b_\alpha$ of opposite statistics.  Hence the
difference between the numbers of bosonic and the number of fermionic
degrees of freedom remains unchanged.

There may be additional problems when the gauge-fixing terms fail to
fix all the gauge degrees of freedom entirely. In that case the ghost
system will have a secondary gauge invariance 
which must be fixed by
repeating the same procedure and introducing a next generation of
ghost fields. Such a phenomenon is known to occur, for instance, for
anti-symmetric tensor gauge fields
\cite{Townsend:1979hd,Siegel:1980jj}. An elegant way to deal with this
situation has been described in \cite{Batalin:1983wj}. Furthermore the
expectation values of the gauge-fixing terms must remain zero at the
quantum level, so that the BRST symmetry will not be realized in a
spontaneously broken way \cite{deWit:1976pn}.

What remains is to consider the extension to the case of a gauge
algebra with both bosonic and fermionic generators. In principle this
extension is standard (see e.g. \cite{Gomis:1994he}), and we briefly
introduce the relevant notation. Let us first consider the matter
fields $\phi^i$, which can refer to either commuting (bosonic) or 
anti-commuting (fermionic) fields. To each field we assign a
statistical index $\epsilon_i$, equal to $0$ when the field is
bosonic and to $1$ when the field is fermionic, so that $\phi^i
\phi^j= (-)^{\epsilon_i\epsilon_j} \phi^j\phi^i$. Likewise we
introduce similar indices $\epsilon_\alpha$ for the transformation
parameters. Note that these indices are defined modulo~2. These
definitions now enable one to define statistical indices for all
quantities involved. For instance we have
\begin{equation}
  \label{eq:index-R-f}
  \epsilon(R^i{\!}_\alpha) \= \epsilon_i+\epsilon_\alpha\,, \qquad
  \epsilon(f_{\alpha\beta}{\!}^\gamma) \=
  \epsilon_\alpha+\epsilon_\beta+\epsilon_\gamma\,, \qquad
  f_{\alpha\beta}{\!}^\gamma \= (-)^{\epsilon_\alpha+\epsilon_\beta}
  f_{\beta\alpha}{\!}^\gamma \,. 
\end{equation}
In the context of BRST the indices of the additional fields and the
parameter follow directly from the definitions above,
\begin{equation}
  \label{eq:index-brst}
  \epsilon(c^\alpha) \=   \epsilon(b_\alpha) \= \epsilon_\alpha + 1 \,,\qquad 
  \epsilon(B_\alpha) \= \epsilon_\alpha \, , \qquad
  \epsilon(\Lambda) \= 1 \, .
\end{equation}
Finally we should point out that the derivative with respect to an
anti-commuting quantity is ambiguous when acting on a commuting
composite. In that case one has to distinguish between a right- and a
left-derivative (whose sum will be vanishing).

Note that in the main body of the paper we assume that all the gauge
field generators are bosonic to avoid heavy notation and to keep the
derivations as clear as possible. This means that, when considering
theories with both bosonic and fermionic generators, one cannot just
copy the results from this paper, because we may have accidentally
ordered the terms in a way that is allowed for the purely bosonic case,
but not for the mixed case.

\section{The background field split} 
\label{sec:back-field}
\setcounter{equation}{0}

As already explained in the introduction we will be dealing with a
gauge theory in the presence of a boundary. At this boundary one must
choose certain boundary conditions and the obvious one is to require
that the boundary will be invariant under a subgroup of the full local
gauge group. Hence one has to distinguish between the transformations
that leave the boundary invariant and the transformations that act in
the bulk, which will be integrated over in the path integral. This can
be done systematically by first performing a background field split
where the background refers to the boundary configuration extended
into the bulk. The quantum fields are then viewed as fluctuations
about this background and will eventually be integrated over in a path
integral. At the boundary the quantum fields will vanish, but for the
moment we refrain from discussing the details of these boundary
conditions. For simplicity we restrict ourselves again to bosonic
fields and transformation parameters.

To set up the background field split, let us consider a gauge theory
with fields generically denoted by $\phi^i$, which are decomposed into
background fields $\mathring{\phi}^{\,i}$ and quantum
fields~$\widetilde{\phi}^{\,i}$.  The latter are the fields that one
has to integrate over in a path integral. This integration requires to
make use of a standard quantization method such as BRST
quantization. The most straightforward decomposition between
background and quantum fields is
\begin{equation}
  \label{eq:bg-split}
  \phi^i \= \mathring\phi{}^{\,i} + \widetilde\phi{}^{\,i}\,, 
\end{equation}
but in specific cases one may prefer to employ more sophisticated
decompositions. Eventually the background fields are fixed at the
boundary of the space and they are continued into the bulk. We assume
that the precise continuation is not important because the deviation
from their value in the bulk is characterized by the quantum fields
which eventually will be integrated out.  The gauge transformations
are as specified in \eqref{eq:gauge-transf} and they can
correspondingly be decomposed in two different ways. The background
transformations $\mathring{\delta}$ take the form,
\begin{equation}
  \label{eq:split-transfo}
  \mathring{\delta}\mathring{\phi}^{\,i} \= R(\mathring{\phi})^i{}_\alpha \,
  \mathring{\xi}^{\,\alpha}\,,  \qquad 
  \mathring{\delta} \widetilde{\phi}^{\,i} \= \Delta
  R(\mathring{\phi}, \widetilde{\phi})^i{}_\alpha \,
  \mathring{\xi}^{\,\alpha}\,, 
\end{equation}
where
$\Delta R(\mathring{\phi}, \widetilde{\phi})^i{}_\alpha \equiv
R(\mathring{\phi}+\widetilde{\phi})^i{}_\alpha
-R(\mathring{\phi})^i{}_\alpha$.
The gauge transformations $\tilde{\delta}$ that are relevant when
integrating over the fields $\widetilde{\phi}^{\,i}$ must leave the
background fields invariant and thus take the form,
\begin{equation}
  \label{eq:true-gauge-tr}
  \tilde{\delta}\mathring{\phi}^{\,i} \=0\,,\qquad
  \tilde{\delta}\widetilde{\phi}^{\,i} \= 
  R(\mathring{\phi}+\widetilde{\phi})^i{}_\alpha \,\xi^\alpha\,,
\end{equation}
and in the following we will keep referring to them as {\it quantum 
transformations}. 
 
We start by considering the commutators of the quantum and background
transformations acting on the background fields. For the background
fields $\mathring{\phi}^{\,i}$ a straightforward calculation yields
\begin{align}
  \label{eq:commm+backgr-on-background}
  \big[\tilde\delta(\xi_1) \, \tilde\delta(\xi_2)
  -\big(1\leftrightarrow2\big)\big] \mathring{\phi}^{\,i} \=&\;0 \,,
  \nonumber\\[2mm] 
  \big[ \mathring{\delta}(\mathring{\xi}) \, \tilde\delta(\xi) -
  \tilde\delta(\xi)\, \mathring{\delta}(\mathring{\xi})\big]
  \mathring{\phi}^{\,i} \=&\; 0 \,, \nonumber\\[2mm]
  \big[ \mathring{\delta}(\mathring{\xi}_1) \,
  \mathring{\delta}(\mathring{\xi}_2)
  -\big(1\leftrightarrow2\big)\big] \mathring{\phi}^{\,i} \=&\;
  f(\mathring{\phi})_{\alpha\beta}{\!}^\gamma \,
  \mathring{\xi}_1{\!}^\alpha\,\mathring{\xi}_2{\!}^\beta \,
  R(\mathring{\phi})^i{\!}_\gamma   \,.
\end{align}
Subsequently one determines the same commutators, but now acting on the
quantum fields, 
\begin{align}
  \label{eq:commm-[backgr+quantum]-on-quantum}
  \big[\tilde\delta(\xi_1) \, \tilde\delta(\xi_2)
  -\big(1\leftrightarrow2\big)\big] \widetilde\phi^{\,i} \=&\;
  f(\mathring{\phi}+\widetilde\phi)_{\alpha\beta}{\!}^\gamma \,
  \xi_1{\!}^\alpha\,\xi_2{\!}^\beta \,
  R(\mathring{\phi}+\widetilde\phi)^i{\!}_\gamma  \,, \nonumber\\[2mm]
  \big[ \mathring{\delta}(\mathring{\xi}) \, \tilde
  \delta(\xi)-    \tilde\delta(\xi)\, \mathring{\delta}(\mathring{\xi})
  \big] \widetilde{\phi}^{\,i} \=&\; 
  f(\mathring{\phi} +\widetilde\phi)_{\alpha\beta}{\!}^\gamma \,
  \mathring{\xi}^\alpha\, {\xi}^\beta \, 
  R(\mathring{\phi} +\widetilde\phi)^i{\!}_\gamma  \,, \nonumber\\[2mm]
     \big[ \mathring{\delta}(\mathring{\xi}_1) \,
      \mathring{\delta}(\mathring{\xi}_2)
      -\big(1\leftrightarrow2\big)\big] \widetilde{\phi}^{\,i} \=&\;
      f(\mathring{\phi})_{\alpha\beta}{\!}^\gamma \,
      \mathring{\xi}_1{\!}^\alpha\, \mathring{\xi}_2{\!}^\beta \, 
      \Delta R(\mathring{\phi},\widetilde{\phi})^i{\!}_\gamma  \nonumber\\
      &\; +  \big[f(\phi)- f(\mathring{\phi})\big]_{\alpha\beta}{\!}^\gamma \,
      \mathring{\xi}_1{\!}^\alpha\, \mathring{\xi}_2{\!}^\beta \;
      R(\mathring{\phi} +\widetilde\phi)^i{\!}_\gamma \,. 
\end{align}
It is clear that the combined quantum and background transformations
generate a closed algebra on $\mathring{\phi}^{\,i}$ and
$\widetilde{\phi}^{\,i}$. Its global structure has the following form,
\begin{equation}
  \label{eq:structure-double-algebra}
  [\,\mathring{\delta}\,, \mathring{\delta}\, ] \= \mathring{\delta} +
  \widetilde{\delta} \,, \qquad 
  [\,\mathring{\delta}\,, \widetilde{\delta}\, ] \=
  \widetilde{\delta} \,, \qquad
    [\,\widetilde{\delta}\,, \widetilde{\delta}\, ] \=
  \widetilde{\delta} \,.
\end{equation}
When the algebra is {\it soft}, meaning that the structure `constants'
depend on the fields, then the background transformation will {\it
  not} form a subgroup. However, the closure of the full algebra 
remains unaffected.

Therefore we can construct a BRST complex by introducing two sets of
ghosts, $\mathring{c}^{\,\alpha}$ and~$c^{\,\alpha}$, corresponding to
the background and the quantum transformations, respectively.  Having
introduced these variables, it is then straightforward to define the
BRST transformations, which will eventually give rise to a nilpotent
BRST charge. The BRST transformation on the fields $\mathring{\phi}^{\,i}$
and $\widetilde\phi^{\,i}$ follows upon substituting
$\xi^\alpha= \Lambda\,c^\alpha$ and
$\mathring{\xi}^\alpha= \Lambda\, \mathring{c}^{\,\alpha}$.  The
result reads as follows,
\begin{align}
  \label{eq:BRST-phi}
  \delta_\mathrm{brst} \mathring{\phi}^{\,i} \=&\,
  R(\mathring{\phi})^i{\!}_\alpha \, \Lambda\, \mathring{c}^{\,\alpha}
                                                 \,, \nonumber\\ 
  \delta_\mathrm{brst} \widetilde\phi^{\,i} \=&\,
  R(\mathring{\phi}+\widetilde{\phi})^i{\!}_\alpha \, \Lambda\,
 (c^\alpha + \mathring{c}^{\,\alpha}) - 
  R(\mathring{\phi})^i{\!}_\alpha \, \Lambda
  \,\mathring{c}^{\,\alpha}   \,.
\end{align}
Here and in the remainder of this paper we will take into account that
the theory contains both commuting and anti-commuting fields and gauge
parameters. As it turns out the corresponding changes are rather
minimal. As before the BRST transformations of the ghost fields follow
straightforwardly from the commutation relations given in
\eqref{eq:commm+backgr-on-background} and~\eqref{eq:commm-[backgr+quantum]-on-quantum} and yield
\begin{align}
  \label{eq:ghost-brst}
  \delta_\mathrm{brst} \, \mathring{c}^{\,\gamma}
  \=&\;\tfrac12\,
  f(\mathring{\phi})_{\alpha\beta}{\!}^\gamma \,
  \mathring{c}^{\,\alpha}\Lambda \,\mathring{c}^{\,\beta} \,,\nonumber\\[2mm]
  \delta_\mathrm{brst} \, c^{\,\gamma} \=&\; \tfrac12\,
  f(\phi)_{\alpha\beta}{\!}^\gamma \, c^\alpha\Lambda \,c^\beta
   + f(\phi)_{\alpha\beta}{\!}^\gamma \,
  \mathring{c}^{\,\alpha}\Lambda \,c^\beta +\tfrac12 \,
  \big[f(\phi)- f(\mathring{\phi})\big]_{\alpha\beta}{\!}^\gamma \,
  \mathring{c}^{\,\alpha}\, \Lambda\, \mathring{c}^{\,\beta}
  \nonumber\\
  \=&\; \tfrac12\, f(\phi)_{\alpha\beta}{\!}^\gamma \,
  (c+\mathring{c})^\alpha\Lambda \,(c+\mathring{c}) ^\beta
  -\tfrac12\, 
  f(\mathring{\phi})_{\alpha\beta}{\!}^\gamma \,
  \mathring{c}^{\,\alpha}\, \Lambda\, \mathring{c}^{\,\beta} \,,
\end{align}
where $\phi^i= \mathring{\phi}^{\,i} +\widetilde{\phi}^{\,i}$. An interesting
observation is that \eqref{eq:ghost-brst} leads to
\begin{equation}
  \label{eq:new-brst-ghost}
  \delta_\mathrm{brst} \, (c+\mathring{c})^{\,\gamma} =
  \tfrac12\,  f(\phi)_{\alpha\beta}{\!}^\gamma \,
  (c+\mathring{c})^\alpha\Lambda \,(c+\mathring{c}) ^\beta
  \,,
\end{equation}
which confirms the consistency of splitting the ghosts into background
ghosts $\mathring{c}^{\,\alpha}$ and quantum ghosts $c^{\,\alpha}$,
even in the case that the gauge algebra is soft! Note that the
anti-ghosts $b_\alpha$ and the Lagrange multiplier fields $B_\alpha$
should be regarded as quantum fields, so that their BRST
transformations remain unchanged and are given by
\begin{equation}
  \label{eq:BRST-Bb}
  \delta_\mathrm{brst}\,b_\alpha \=  \Lambda\, B_\alpha \,, \qquad 
  \delta_\mathrm{brst}\,B_\alpha \= 0\,.
\end{equation}

The closure of the underlying gauge algebra expressed by the closure
relations \eqref{eq:commm+backgr-on-background} and
\eqref{eq:commm-[backgr+quantum]-on-quantum} now guarantees that the
BRST charge is nilpotent, which can also be verified by explicit
calculation, 
\begin{equation}
  \label{eq:nilpotency}
  \delta_\mathrm{brst}{\!}^2 \=0\,. 
\end{equation}

The corresponding BRST invariant action is a generalization of
\eqref{eq:BRST-Lagrangian}. However, in this case one introduces
only anti-ghosts $b_\alpha$ and Lagrange multipliers $B_\alpha$
associated with the quantum fields; for the background fields there
will be no gauge-fixing terms. The quantum action then takes
the form,
\begin{align}
  \label{eq:BRST-action}
    S_\mathrm{brst}[\widetilde{\phi}^{\,i}, c^\alpha, b_\alpha, B_\alpha; 
  \mathring{\phi}^{\,i}, \mathring{c}^{\,\alpha}] 
   \=&  \int
  \mathrm{d}^{n} x \,\Big[ 
 \mathcal{L}^\mathrm{class}(\mathring{\phi}+\widetilde{\phi} ) + B_\alpha 
  \, F(\mathring{\phi},\widetilde{\phi})^\alpha  \\
  &\quad\qquad
  - (-)^{\epsilon_\alpha+\epsilon_\beta +\epsilon_j}\,
     b_\alpha \,R(\mathring{\phi} +\widetilde{\phi})^j{\!}_\beta
   \,(c +\mathring{c})^{\,\beta}
  \,\tilde\partial_j F(\mathring{\phi}, \widetilde{\phi} )^\alpha \nonumber\\
  &\quad\qquad  
    - (-)^{\epsilon_\alpha+\epsilon_\beta +\epsilon_j}\,
     b_\alpha \,R(\mathring{\phi} )^j{\!}_\beta
   \,\mathring{c}^{\,\beta}
  \,(\mathring{\partial} -\tilde{\partial})_j F(\mathring{\phi}, 
    \widetilde{\phi} )^\alpha   \Big] \,. \nonumber 
\end{align}
With suitable boundary conditions this is a BRST invariant functional
of both the quantum and the background fields. Here we have assumed
that the fields live in an $n$-dimensional space, and
$\mathring{\partial}_j$ and $\widetilde{\partial}_j$ denote the
derivatives with respect to $\mathring{\phi}^{\,j}$ and
$\widetilde{\phi}^{\,j}$, respectively. In the above equation they are
defined as left-derivatives. Furthermore the gauge conditions
$F^\alpha$ should be non-singular, meaning that $F(\mathring{\phi},
\widetilde{\phi} )^\alpha=0$ must fix the values of the quantum fields
$\widetilde{\phi}^{\,i}$. Finally we observe that the ghosts $c^{\,\alpha}$
and $\mathring{c}^{\,\alpha}$ carry ghost number $+1$, whereas the
anti-ghosts $b_\alpha$ carry ghost number $-1$, so that the action
\eqref{eq:BRST-action} carries zero ghost number.

The next step is to consider a functional integral over the quantum
fields $\widetilde{\phi}^{\,i}$ and $c^{\,\alpha}$, $b_\alpha$ and
$B_\alpha$, 
\begin{equation}
  \label{eq:Z}
  Z[\mathring{\phi}] \= \int \,\mathcal{D}\widetilde{\phi}^{\,i}\,
  \mathcal{D}c^\alpha \, \mathcal{D}b_\alpha\, 
  \mathcal{D}B_\alpha \;  \exp\Big[  S_\mathrm{brst}[\widetilde{\phi}^{\,i},
  c^\alpha, b_\alpha, B_\alpha;  
  \mathring{\phi}^{\,i}, \mathring{c}^{\,\alpha}]  \Big] \,.  
\end{equation}
One can show that the restricted functional integration measure is BRST
invariant under the same conditions as the full functional integral
without background field splitting, namely
\begin{align}
  \label{eq:measure}
  \partial_i   R(\phi)^i{\!}_\alpha \=0 \,, \qquad
  f(\phi)_{\alpha\beta}{\!}^\beta \=0\,.  
\end{align}
Since the indices on the fields include their space-time arguments,
these two expressions are proportional to $\delta^n(0)$, where
$\delta^n(x)$ is an $n$-dimensional delta function. Consequently they are 
ill-defined. This is a known complication, which has
been studied in the past (see, for instance,
\cite{Fujikawa:1987ie,Bern:1990bh}). On the basis of that we will
assume from now on that the path integral in \eqref{eq:Z} is indeed
fully consistent with regard to BRST transformations.  Note that the
action may still contain additional terms that depend exclusively on
$B_\alpha$, because this field is BRST invariant. Irrespective of
this, the integration measure for the fields $b_\alpha$ and $B_\alpha$
is BRST invariant by itself, so that no further modifications are
required.

As already anticipated in the notation, the path integral
$Z[\mathring{\phi}]$ does not depend on the background ghosts. This
follows directly from the observation that the right-hand side carries
zero ghost number. Indeed, one can easily verify that the terms in
\eqref{eq:BRST-action} that are proportional to
$b_\alpha \, \mathring{c}{}^{\,\beta}$ will not contribute to the
functional integral.  We have thus established that
\begin{equation}
  \label{eq:brst-Z-0}
  \delta_\mathrm{brst} Z[\mathring{\phi}] =  \frac{\partial Z[\mathring{\phi}]}
  {\partial\mathring{\phi}^{\,i}}  \, R(\mathring{\phi})^i{\!}_\alpha\, \Lambda \,
  \mathring{c}^{\,\alpha}\,,  
\end{equation}
so that the functional integral is fully BRST invariant when the
background specified by the fields $\mathring{\phi}^{\,i}$ is
invariant. Clearly the background ghosts only play an ancillary role
here as the parameters that specify the background transformations.
The existence of a consistent BRST complex that involves both quantum
and background fields with corresponding ghost fields is a non-trivial
result. It is remarkable that this result also holds for theories with
a soft gauge algebra, where the structure constants depend on the
fields.
 
To prove that the path integral \eqref{eq:Z} does not depend on the
gauge condition, we first extend it by including external sources
coupling to single fields or to composite operators. In this way one
obtains a generating functional for Green's functions in a particular
gauge, which can be used to derive BRST Ward identities. Hence we
include an exponential factor with a variety of external sources into
the integrand of the path integral \eqref{eq:Z},\footnote{
  External sources coupling to background fields are not revelvant
  here as the path integral does not involve an integration over these
  fields. } 
\begin{equation}
  \label{eq:ext-sources}
  \exp \int \mathrm{d}^n x\, \big[ J_b{\!}^\alpha(x)\, b_\alpha(x) +
  \widetilde{J}_i(x)\,\widetilde{\phi}^{\,i}(x)  + J^c{\!}_\alpha(x) \, c^\alpha(x)
  +  {J}_B{\!}^\alpha(x)\,B_\alpha(x) + \cdots \big]   \,.
\end{equation}
The expansion of the path integral in terms of the external sources
defines the corresponding Green's functions.  Shifting the fields in
\eqref{eq:ext-sources} by the BRST-transformed fields leads to a
rearrangement of Green's functions, while, on the other hand, the extra
terms can be eliminated by making use of the fact that
$S_\mathrm{brst}$ and the integration measure of the functional
integral is BRST invariant, up to the transformations of the
background fields. In this way one thus obtains the Ward identities
between Green's functions.  There is an implicit assumption here,
namely that BRST symmetry is manifest and not realized in a
spontaneously broken way. If that were not the case, then the
invariant action would contribute to the Ward identities in the form
of the divergence of the BRST Noether current.

Let us now derive two Ward identities and discuss their
consequences. In the first one we put all sources to zero with the
exception of $J_b{\!}^\alpha$. The Ward identity then takes the form
\begin{align}
  \label{eq:WI-1}
  &\int \,\mathcal{D}\widetilde{\phi}^{\,i}\,
  \mathcal{D}c^\alpha \, \mathcal{D}b_\alpha\, 
  \mathcal{D}B_\alpha \;  \exp\Big[  S_\mathrm{brst}  + \int
  \mathrm{d}^n y \,  J_b{\!}^\gamma(y)\, b_\gamma(y)  
 \Big] \,\int \mathrm{d}^nx \,J_b{\!}^\alpha(x)\, \Lambda B_\alpha(x) =0\,,
\end{align}
where we used the BRST variation of the anti-ghost field.  Only the
term linear in $J_b{\!}^\alpha$ can give a non-zero contribution,
because the ghost fields in the action are all paired with anti-ghost
fields. Since the source is not subject to any restriction it thus
follows that the expectation value of $B_\alpha$ must vanish, i.e.
\begin{align}
  \label{eq:WI-1}
  &\int \,\mathcal{D}\widetilde{\phi}^{\,i}\,
    \mathcal{D}c^\alpha \, \mathcal{D}b_\alpha\, 
    \mathcal{D}B_\alpha \;  \exp\big[  S_\mathrm{brst}\big]\;  B_\alpha(x) \=0\,.
\end{align}
On the other hand, whether or not the expectation value of $B_\alpha$
will vanish is eventually a dynamical question that depends on the
details of the action $S_\mathrm{brst}$. When the expectation value does not
vanish, the BRST symmetry will be realized in a spontaneously broken
way in view of the fact that the expectation value of
$\delta_\mathrm{brst} b_\alpha$ will not vanish. In that case the Ward
identity will receive extra contributions as we already indicated
previously. However, it is obvious that this option is of no physical
interest, and one has to insist that BRST invariance is manifestly
realized~\cite{deWit:1976pn}.

For the second Ward identity we keep the dependence on the source
$J_b{\!}^\alpha$ but in addition we now consider a second source
coupling to a composite operator
$\Delta F(\widetilde{\phi}, \mathring{\phi})^\beta$. The terms of
higher order in $J_b{\!}^\alpha$ will not contribute, just as in the
previous case, and we will restrict ourselves to the first-order
contribution in the composite operator. By differentiating with
respect to the two external sources one thus derives the following
Ward identity,
\begin{align}
  \label{eq:2}
  &\int \,\mathcal{D}\widetilde{\phi}^{\,i}\,
    \mathcal{D}c^\alpha \, \mathcal{D}b_\alpha\, 
    \mathcal{D}B_\alpha \;  \exp\big[ S_\mathrm{brst} \big]  \nonumber \\ 
  &\times \Big[  \Delta F(\widetilde{\phi},
    \mathring{\phi})^\beta(y)\,   \Lambda B_\alpha (x) +   \delta_\mathrm{brst}
    \Delta F(\widetilde{\phi}, 
    \mathring{\phi})^\beta(y)\,b_\alpha(x)  \Big] \= 0. 
\end{align}
Upon integrating this result over $x$ and $y$ with a delta function
$\delta^n(x-y)$ and contracting the indices with
$\delta^\alpha{\!}_\beta$, one recognizes that this result is precisely
the original result \eqref{eq:Z} for $Z[\mathring{\phi}]$ but now with
a gauge-fixing term equal to
$F(\widetilde{\phi},\mathring{\phi})^\alpha +\Delta
F(\widetilde{\phi},\mathring{\phi})^\alpha$,
expanded to first order in $\Delta F^\alpha$. This proves that
$Z[\mathring{\phi}]$ is independent of the choice of the gauge
condition.\footnote{
This particular argument is a slight generalization of the
analysis presented in \cite{tHooft:1972qbu}, which was used to derive
the gauge independence of the S-matrix in gauge theories with
quadratic gauge fixing (where BRST is not nilpotent on the anti-ghost
fields $b_\alpha$).} 

An interesting observation in view of what will be discussed later, is
that the gauge independence is not affected when including extra terms
in the action that are BRST exact,~i.e. terms that can be written as
the BRST variation of functions of the fields $\widetilde{\phi}^{\,i}$ and
$\mathring{\phi}^{\,i}$. In the specific context of BRST quantization this
observation is not particularly useful, as these terms will violate
ghost number conservation. Only the gauge-fixing term, which is also
BRST exact, will preserve ghost number by virtue of the presence of the
anti-ghost field.

\section{Equivariant cohomology}
\label{sec:equiv-cohom}
\setcounter{equation}{0}

In the previous section we have presented a consistent background
field split in which the original fields have been decomposed into
background and quantum fields, denoted by $\mathring{\phi}^{\,i}$ and
$\widetilde{\phi}^{\,i}$, respectively, thus doubling all the
fields. Correspondingly we have also doubled the gauge transformations
in terms of background and quantum gauge transformations, and we have
shown that they can be incorporated consistently into an extended BRST
complex.  This extension can be given irrespective of whether the 
gauge algebra is soft or not.  We only used that the gauge
transformations close off shell.

This particular set-up was proposed in order to deal with gauge
theories in the presence of a boundary. The boundary values of the
original fields, which will be motivated primarily by physical
considerations, are carried by the background fields
$\mathring{\phi}^{\,i}$ that will be smoothly continued into the
bulk. The quantum fields $\widetilde{\phi}^{\,i}$, on the other hand,
describe the fluctuations in the bulk about the selected background
fields; obviously the quantum fields must therefore vanish at the
boundary. Their fluctuations will eventually be integrated over in a
suitable path integral as was shown in the previous section.

The background fields $\mathring{\phi}^{\,i}$ 
will typically be invariant under an isometry
group that is a subgroup of the full group of background
transformations. In the continuation of the background fields into the
bulk, the isometry group has to remain manifest. The background ghosts
should then be restricted to take their values in the isometry algebra. All this
implies that the BRST transformations on the background fields are
necessarily constrained to vanish,
\begin{equation}
  \label{eq:isometry}
  \delta_\mathrm{brst} \mathring{\phi}^{\,i} \=
    R(\mathring{\phi})^i{\!}_\alpha\,\Lambda\, \mathring{c}^{\,\alpha} \= 0\,.    
\end{equation} 
Consequently the background ghosts $\mathring{c}^{\,\alpha}$, which
play the role of symmetry parameters associated to the background
transformations, should vanish with the exception of those that
parametrize the isometry group of the background field
configuration. Since the isometry group is a subgroup of the full
background symmetry group, this ensures that the above restriction can
be imposed consistently. Here we are implicitly assuming that the
isometry group is defined for the global background field
configuration (i.e.~also in the bulk), which poses a restriction on
how the background fields are continued into the bulk. The
non-vanishing background ghosts $\mathring{c}^{\,\alpha}$ that
parametrize the isometry group will in general be subject to
differential constraints that are implied by the appropriate Killing
equations associated with the background isometries.  Under these
conditions the structure constants of the background isometry algebra
follow obviously from the original structure constants
$f(\mathring{\phi})_{\alpha\beta}{\!}^\gamma$ upon considering the
explicit embedding of the isometry group into the full background
symmetry group. As far as the BRST transformations are concerned the
possible field-dependence of
$f(\mathring{\phi})_{\alpha\beta}{\!}^\gamma$ is not relevant in view
of the constraint \eqref{eq:isometry}. Because of this constraint the
BRST transformations of the quantum fields $\widetilde\phi^{\,i}$
simplify and take the form
\begin{equation}
  \label{eq:truncate-brst-quantum}
  \delta_\mathrm{brst} \widetilde\phi^{\,i} \= 
  R(\mathring{\phi}+\widetilde{\phi})^i{\!}_\alpha \, \Lambda\,
 ( c +\mathring{c})^{\alpha}  \,.
\end{equation}
Note that one can subsequently consider a possible subalgebra of the
isometry algebra by further restricting the number of background
ghosts. In the subsequent discussion it will be important that some of
the background ghosts remain present and will generate a non-trivial
subgroup of the background isometries, so that \eqref{eq:isometry}
remains valid.

Let us now continue and consider the BRST transformation on the
background ghosts,
\begin{equation}
  \label{eq:truncate-brst-c0}
     \delta_\mathrm{brst} \mathring{c}^{\,\alpha} \= \tfrac12\,
  f(\mathring{\phi})_{\beta\gamma}{\!}^\alpha \,
  \mathring{c}^{\,\beta}\Lambda \,\mathring{c}^{\,\gamma} \,. 
\end{equation}
This variation is consistent with the reduction of the background
ghosts to the isometry algebra. Therefore there is no need for
introducing any additional constraints on
$f(\mathring{\phi})_{\alpha\beta}{\!}^\gamma \,
\mathring{c}^{\,\alpha}\Lambda \,\mathring{c}^{\,\beta}$\,.  As a
result the BRST transformations on the quantum ghosts remain
unchanged,
\begin{equation}
  \label{eq:BRST-c}
  \delta_\mathrm{brst} \, c^{\,\alpha} \=  
  \tfrac12\, f(\phi)_{\beta\gamma}{\!}^\alpha \,
  (c+\mathring{c})^\beta\Lambda \,(c+\mathring{c})^\gamma
  -\tfrac12\, 
  f(\mathring{\phi})_{\beta\gamma}{\!}^\alpha \,
  \mathring{c}^{\,\beta}\, \Lambda\, \mathring{c}^{\,\gamma}  \,.
\end{equation} 
It is now straightforward to verify that the BRST symmetry is still
nilpotent. As before this requires to use the Jacobi identity
\eqref{eq:Jacobi}, which simplifies for the background structure
constants because $\mathring{\phi}^{\,i}$ is now BRST
invariant. Furthermore it follows from \eqref{eq:brst-Z-0} that the
path integral \eqref{eq:Z} is BRST invariant as well. 
 
None of the quantum fields are constrained, and therefore they will
appear as before in the functional integral~\eqref{eq:Z}; this
integral now involves a coupling to a restricted set of background
fields, $\mathring{\phi}^{\,i}$ and $\mathring{c}^{\,\alpha}$, but
nevertheless it remains well-defined, also in view of the fact that
the functional integral did not include an integration over the
background fields and ghosts.

Until now we did not change the original BRST algebra, but rather we
adopted a special field representation by requiring that the
background fields $\mathring{\phi}^{\,i}$ were BRST invariant. This
implied that the background ghosts $\mathring{c}^{\,\alpha}$ had to be
restricted to take their values in the corresponding isometry
subalgebra.  As a next step we now introduce a {\it deformation} of
the BRST algebra by imposing the condition that also background ghosts
will remain invariant under the algebra, without implying that the
right-hand side of \eqref{eq:truncate-brst-c0} must vanish. Upon
imposing this deformation both the background fields and the
background ghosts will thus remain invariant, while the
transformations of the quantum fields are unchanged.  We denote the
resulting variations by $\delta_\mathrm{eq}$, which take the following
form,
\begin{align}
  \label{eq:equivariant-tr}
     \delta_\mathrm{eq}\,\mathring{\phi}^{\,i} \= &\;0  \,, 
   \qquad  \quad 
   \delta_\mathrm{eq} \, \mathring{c}^{\,\alpha} \= 0 \,, \nonumber\\
  \delta_\mathrm{eq}\,\widetilde\phi^{\,i} \=& \;
  R(\mathring{\phi}+\widetilde{\phi})^i{\!}_\alpha \, \Lambda\,
 ( c^\alpha +\mathring{c}^{\,\alpha}) \,, \nonumber\\
  \delta_\mathrm{eq} \, c^{\,\alpha} \=& \;
  \tfrac12\, f(\phi)_{\beta\gamma}{\!}^\alpha \,
  (c+\mathring{c})^\beta\Lambda \,(c+\mathring{c})^\gamma
  -\tfrac12\, 
  f(\mathring{\phi})_{\beta\gamma}{\!}^\alpha \,
  \mathring{c}^{\,\beta}\, \Lambda\, \mathring{c}^{\,\gamma}
  \,.
\end{align}
As the reader can verify these transformations are no longer
nilpotent. Instead they define an {\it equivariant map}.  The relevant
relations, which follow again by making use of the closure relation
\eqref{eq:closure} and the Jacobi identity~\eqref{eq:Jacobi}, are
\begin{equation}
  \label{eq:equivariant}
  \delta_\mathrm{eq}{}^2 \= \delta_{\mathring{\xi}}\,, \qquad \qquad 
  [\delta_\mathrm{eq}\,, \delta_{\mathring{\xi}}] \= 0\,.
\end{equation}
The new transformation $\delta_{\mathring{\xi}}$ acts on the
quantum fields according to
\begin{align}
  \label{eq:def-delt-xi}
  \delta_{\mathring{\xi}}\, \widetilde{\phi}^{\,i} \=&\;    
   R(\mathring{\phi}+\widetilde{\phi})^i{\!}_\alpha
  \,\mathring{\xi}^\alpha \,, \nonumber\\
  \delta_{\mathring{\xi}}\, c{}^{\,\alpha} \=&\;
   f(\mathring{\phi}+\widetilde{\phi})_{\beta\gamma}{\!}^\alpha\, 
  (c+\mathring{c})^\beta\,\mathring{\xi}^\gamma \,,
\end{align}
with the transformation parameter $\mathring{\xi}^\alpha$
equal to 
\begin{equation}
  \label{eq:def-xi-0}
  \mathring{\xi}^\alpha \; \equiv \; \Lambda_{[2} \,
  f(\mathring{\phi})_{\beta\gamma}{\!}^\alpha \, 
  \mathring{c}^{\,\beta}\, \Lambda_{1]}\, \mathring{c}^{\,\gamma}
  \, .
\end{equation}
Note that the $\mathring{\xi}^\alpha$ take their values in the
isometry algebra. The background fields and ghosts are obviously
invariant under $\delta_{\mathring{\xi}}$.

The equivariant algebra~\eqref{eq:equivariant} must also be defined on
the anti-ghosts and the Lagrange multiplier fields. Assuming that
$\delta_\mathrm{eq}\,b_\alpha$ coincides with
$\delta_\mathrm{brst}\,b_\alpha$, one deduces the form of
$\delta_\mathrm{eq} \,B_\alpha$,
\begin{align}
  \label{eq:eq-bB}
    \delta_\mathrm{eq} \, b_\alpha\=& \; \Lambda\, B_\alpha \,,
    \nonumber\\
  \delta_\mathrm{eq} \, B_\alpha \=& \; \tfrac12\,
   f(\mathring{\phi})_{\delta\varepsilon}{\!}^\beta \,  
  \mathring{c}^{\,\delta} \,\Lambda \,\mathring{c}^{\,\varepsilon} \, 
  f(\mathring{\phi})_{\alpha\beta}{\!}^\gamma \,b_\gamma   \,.
\end{align}
The action of
$\delta_{\mathring{\xi}}$ on both $b_\alpha$ and $B_\alpha$ then
follows from imposing the algebra \eqref{eq:equivariant}. The result is
\begin{align}
  \label{eq:xi-bB}
  \delta_{\mathring{\xi}}\, b_\alpha \=&\;  \mathring{\xi}^\beta\,
    f(\mathring{\phi})_{\alpha\beta}{\!}^\gamma\,b_\gamma \,,
    \nonumber\\
  \delta_{\mathring{\xi}}\,B_\alpha \=&\; \mathring{\xi}^\beta\,
    f(\mathring{\phi})_{\alpha\beta}{\!}^\gamma\,B_\gamma\, .
\end{align}
The variations~$\qeq$ defined in~\eqref{eq:equivariant-tr},
\eqref{eq:eq-bB}, and $\delta_{\mathring{\xi}}$ defined in
\eqref{eq:def-delt-xi}, \eqref{eq:xi-bB}, have a well-defined ghost
number equal to 1 and 2, respectively. 

One expects that the boundary should be invariant under both
$\delta_\mathrm{eq}$ and $\delta_{\mathring{\xi}}$.  This is directly
confirmed by applying the generators of the equivariant algebra on the
quantum fields $\widetilde\phi^{\,i}$, $c^\alpha$, $b_\alpha$, and
$B_\alpha$, which themselves vanish at the boundary. Indeed it is easy
to verify that their variations under $\delta_\mathrm{eq}$ and
$\delta_{\mathring{\xi}}$ vanish also at the boundary by virtue of
\eqref{eq:isometry} and the Jacobi 
identity for the structure constants of the background isometry
algebra. Note that this is a local result. The global boundary can
only be invariant provided it contains no singular points. Especially
for spaces of Minkowskian signature this may be an issue. 
Here we will ignore this subtlety and assume that the boundary is
indeed regular. 

The above considerations provide us with a special background 
isometry~$\qeq$ that obeys~$\qeq{\!}^2 = \delta_{\mathring{\xi}}$ 
and acts on all the quantum
fields while leaving the background fields and ghosts invariant. Hence
the quantum fields do transform under the isometries of the background
and their transformation rules are specified by the terms in
$\delta_\mathrm{eq}$ proportional to the background ghosts
$\mathring{c}^{\,\alpha}$.\footnote{
  Alternative ways of modifying the BRST algebra have been described in
  the literature (see e.g.~\cite{Baulieu:2012jj,Bae:2015eoa,
    Costello:2016mgj,Imbimbo:2018duh}), but they are conceptually
  different from the present proposal.} 

We already concluded that the functional integral~$Z[\mathring{\phi}]$
in~\eqref{eq:Z} is a BRST invariant functional of the background
fields~$\mathring{\phi}^{\,i}$, so that the BRST invariance of
$Z[\mathring{\phi}]$ seems to imply its invariance under
$\delta_\mathrm{eq}$, and therefore also under
$\delta_{\mathring{\xi}}$. This expectation is indeed confirmed by
explicit computations.  According to \eqref{eq:equivariant-tr} and
\eqref{eq:eq-bB} the operator~$\delta_\mathrm{eq}$ differs from its
nilpotent ancestor $\delta_\mathrm{brst}$ only in its action on the
background ghosts~$\mathring{c}^{\, \a}$ and the Lagrange multiplier
fields $B_\alpha$.  Bearing in mind that $\mathring{\phi}^{\,i}$ is
invariant, $\delta_\mathrm{eq}(\mathring{\phi} +\widetilde{\phi})^i$
is identical to the original BRST transformation so that the classical
Lagrangian $\CL^\text{class}$ is also invariant under~$\delta_\mathrm{eq}$. 
However, the gauge-fixing term $\CL^\text{g.f.}$
does explicitly depend on $\mathring{c}^{\,\alpha}$ and $B_\alpha$, so
let us us take a closer look. First we note that the last line present
in \eqref{eq:BRST-action} will now vanish by virtue of
\eqref{eq:isometry}. Therefore the gauge-fixing term that appears in
\eqref{eq:BRST-action} becomes identical to (for convenience we
specialize again to commuting gauge transformations and commuting
fields $\mathring{\phi}^{\,i}$, $\widetilde{\phi}^{\,i}$),
\begin{equation}
  \label{eq:gauge-fixing}
  \mathcal{L}^\mathrm{g.f.} = B_\alpha\, F(\mathring{\phi},
  \widetilde{\phi})^\alpha - b_\alpha\,\delta_\mathrm{eq} F(\mathring{\phi},
  \widetilde{\phi})^\alpha\,. 
\end{equation}
It is now clear that $\qeq \CL^\text{g.f.}$ does not vanish. Instead
it will be proportional to $b_\alpha$, resulting from the variation of
$B_\alpha$ given in \eqref{eq:eq-bB} and from the fact that
$\delta_\mathrm{eq}{\!}^2F^\alpha$ is non-vanishing and equal to
$\delta_{\mathring{\xi}} \,F^\alpha$. Not surprisingly these terms
combine into the $\delta_{\mathring{\xi}}$ variation of the gauge fermion
$b_\alpha\,F(\mathring{\phi},\widetilde{\phi})^\alpha$. Therefore we
conclude that the action
\begin{equation} 
  \label{eq:Seq}
  S_\mathrm{eq}[\widetilde{\phi}^{\,i}, c^\alpha, b_\alpha, B_\alpha; 
  \mathring{\phi}^{\,i}, \mathring{c}^{\,\alpha}] 
  \= \int
  \mathrm{d}^{n} x \,\Big[ 
 \mathcal{L}^\mathrm{class}(\mathring{\phi}+\widetilde{\phi} ) + 
 \partial_\Lambda\,\delta_\mathrm{eq}
  \big[ b_\alpha\, F(\mathring{\phi},\widetilde{\phi})^\alpha\big] \Big]
\end{equation}
satisfies 
\begin{equation} 
  \label{eq:Seq-inv}
  \delta_\mathrm{eq}  \, S_\mathrm{eq}  \=  \delta_{\mathring{\xi}}
  \int  \mathrm{d}^{n} x \,   \big[ b_\alpha\,
  F(\mathring{\phi},\widetilde{\phi} )^\alpha\big] \, ,
\end{equation}
where we wrote the variation $\delta_{\mathring{\xi}}$ outside the
integral in view of the fact that the boundary is invariant. 

The functional integral $Z[\mathring{\phi}]$ can now also be written
as
\begin{equation}
  \label{eq:Zeq}
  Z[\mathring{\phi}] \= \int \,\mathcal{D}\widetilde{\phi}^{\,i}\,
  \mathcal{D}c^\alpha \, \mathcal{D}b_\alpha\, 
  \mathcal{D}B_\alpha \;  \exp\Big[  S_\mathrm{eq}[\widetilde{\phi}^{\,i},
  c^\alpha, b_\alpha, B_\alpha;  
  \mathring{\phi}^{\,i}, \mathring{c}^{\,\alpha}]  \Big] \,,
\end{equation}
because the right-hand side of \eqref{eq:Seq-inv} will cancel under
the functional integral over the ghost fields for the simple reason
that it generates terms proportional to the anti-ghosts
without corresponding ghosts. Furthermore the functional
integration measure is also invariant under~$\qeq$ since the
contributions of the variation from the ghosts and quantum fields
vanish by our earlier assumptions \eqref{eq:measure}, and the
transformations of the anti-ghosts~$b_\alpha$ and Lagrange
multipliers~$B_\alpha$ have a trivial Jacobian.  Putting these facts
together, we reach the conclusion that indeed~$\qeq$ is a symmetry of
the functional integral~\eqref{eq:Z}, i.e.,
\begin{equation}
  \label{eq:Z-eq-inv}
  \delta_\mathrm{eq} \, Z[\mathring{\phi}] \= 0 \, .  
\end{equation}
Although there was no need for requiring that \eqref{eq:Seq-inv} must
vanish in order to prove that $Z[\mathring{\phi}]$ is invariant under
$\delta_\mathrm{eq}$, we should point out that the situation will be
different when considering deformations of the integrand. Therefore we
will assume henceforth that the background ghosts are chosen 
so that the background isometry
$\delta_{\mathring{\xi}}$ is compact, so that integrals as in
\eqref{eq:Seq-inv} will generically vanish. 

Finally we consider the dependence of the functional integral on the
gauge condition. As it turns out, one can use the same strategy as
followed at the end of Section~\ref{sec:back-field} to show that the
functional integral is gauge independent. One can also verify that
deformations of the functional integral associated with
$\delta_\mathrm{eq}$-exact terms will leave the gauge independence
unaffected, provided that $\delta_{\mathring{\xi}}$ is compact. 
In this respect, the situation is similar to that of the BRST complex, 
discussed in Section \ref{sec:back-field}.

\section{Localization of the functional integral}
\label{sec:localization}
\setcounter{equation}{0}

The formulation developed in the previous sections seems ideally
suited for applying localization in a large class of theories that
admit local supersymmetry transformations as part of their gauge
algebra.  In particular, we are now able to generalize previous
applications of localization, which so far have mainly been confined
to gauge theories with rigid supersymmetry, to theories of
supergravity.  To do so, consider the functional integral
\eqref{eq:Zeq} where~$\mathcal{L}^\mathrm{class}$ is a supergravity
Lagrangian.\footnote{\label{foot:formal}
  The functional integral in quantum field theory is, of course, only a
  formal physical concept that is not well-defined, especially not in
  quantum gravity, because of severe short-distance singularities. As
  in many supersymmetric theories, the hope is that supersymmetry
  holds at all scales, and that the formal procedure based on
  localization will be valid, irrespective of the serious
  complications in the perturbative context.
} 
The formalism does not rely on the particular form of the classical
Lagrangian, and we are able to discuss supergravity theories which
also include higher-derivative couplings, such as those discussed in
\cite{LopesCardoso:2000qm,deWit:2010za,Butter:2013lta}.  Observe that
in an off-shell formulation of supergravity, the gauge-fixing
described in Sections~\ref{sec:brst-general} and~\ref{sec:back-field}
results in an equal number of bosonic and fermionic degrees of
freedom. The manipulations described in Section~\ref{sec:equiv-cohom}
will only affect the number of background fields and ghosts, so that
the quantum fields will still comprise an equal number of bosonic
and fermionic fields. This is a useful feature of the covariantly
quantized off-shell theory that we will use below. Concerning the
background fields and ghosts, we assume that the background isometries
constitute a rigid superalgebra. The invariance under these isometries
then allows one to consider a purely bosonic background.

Let us now turn to the localization strategy for evaluating
\eqref{eq:Zeq}.  The main idea is to deform the functional integral to
reach a convenient point in field space where we can evaluate it exactly
by using semiclassical methods. Such a deformation
$Z[\mathring{\phi}] = Z[\mathring{\phi};0] \to
Z[\mathring{\phi};\lambda]$
is defined by a corresponding deformation of the action
$S_\mathrm{eq}$ given in \eqref{eq:Seq}, by
$S_\mathrm{eq} = S(0) \to S(\lambda) = S(0) + \lambda \, \qeq\CV$,
where $\lambda$ is a real deformation parameter. The expression for
$\CV$ is chosen to satisfy $\qeq{\!}^2\,\CV =0$, so that the
deformation is $\delta_\mathrm{eq}$-exact and
$\delta_{\mathring{\xi}}\mathcal{V}=0$.  Differentiating with respect
to the parameter $\lambda$ pulls down a factor of~$\qeq\CV$ in the
functional integral, so that we can write
\begin{equation}
  \label{eq:ddtZ} 
  \frac{d}{d\lambda} \,Z[\mathring{\phi};\lambda] \= \int
  \,\mathcal{D}\widetilde{\phi}^{\,i}\, \mathcal{D}c^\alpha \,
  \mathcal{D}b_\alpha\, \mathcal{D}B_\alpha \; \qeq \,\big[
  \mathcal{V}\, \exp[  S_\mathrm{eq} + \lambda \, \qeq \CV]\,\big] \,.  
\end{equation} 
Here we have used that $S_\mathrm{eq}$ vanishes under the action of
$\delta_\mathrm{eq}$, based on the restriction that the background
isometry $\delta_{\mathring{\xi}}$ should be compact (see the comment
at the end of Section~\ref{sec:equiv-cohom}). 

Assuming that $\qeq$ can be represented as a differential operator in
field space~\cite{Schwarz:1995dg}, we conclude that
\begin{equation}
  \label{eq:ddtZ-vanish}
   \frac{d}{d\lambda}\, Z[\mathring{\phi};\lambda] \= 0\,. 
\end{equation}
It is important to mention that one of the conditions for localization 
is that the manifold on which the theory is defined is compact, 
which can only be
achieved in the situations we will be considering by introducing a
cut-off on the asymptotics, as is for instance done in
$\mathrm{AdS/CFT}$ calculations. Our formalism enables us to consider
such a boundary in a systematic way that is consistent with
supersymmetry, but one still has to investigate whether sending the
cut-off to infinity will not introduce any undesirable
effects. Assuming that this is not the case, then
$Z[\mathring{\phi};\lambda]$ will be independent of the deformation
parameter.

An immediate consequence of the property~\eqref{eq:ddtZ-vanish} is
that we can evaluate the original functional integral by taking the
parameter~$\lambda$ to be very large in order to reach a convenient
point in field space.  In this regime,~$Z[\mathring{\phi};\lambda]$
\emph{localizes} to the critical points of the
deformation~$\qeq\mathcal{V}$.  To explain how this limit works in
detail, we make a convenient choice for the deformation by adopting
the following definition\footnote{
  Here the bar on the fermions $\psi$ indicates an appropriate
  conjugation. The action of this conjugation on the fields of the
  theory is known to be subtle even in gauge theories with rigid
  supersymmetry, as there is always some tension between the reality
  conditions of fields and positive-definiteness of $\qeq \CV$. The
  recent work~\cite{deWit:2017cle} on Euclidean supergravity may help
  in clarifying this issue.} 
\begin{equation}
  \label{eq:quadr-deformation}
  \mathcal{V}  \= \int
  \mathrm{d}^nx\,
  \sum_{\bar\imath} \sqrt{\mathring{g}} \;   \overline{\psi}{}_{\bar\imath} \; \delta_\mathrm{eq}
  \psi^{\bar\imath} \, ,
\end{equation}
where we have introduced a suitably chosen background space-time
metric and the sum involves all the fermion fields belonging to the
quantum fields $\widetilde{\phi}^{\,i}$. Correspondingly, we have
split the index $i$ into bosonic and fermionic indices denoted
by~$\hat\imath$ and~$\bar\imath$, respectively.
We remind the reader that we have previously imposed the
condition that $\qeq{\!}^2\, \CV$ must vanish, so that
\begin{equation}
  \label{eq:delta-xi-V}
  \delta_{\mathring{\xi}} \mathcal{V} \= \delta_{\mathring{\xi}}
  \int  \mathrm{d}^{n} x \,
  \sum_{\bar\imath} \, \sqrt{\mathring{g}} \;
  \overline{\psi}_{\bar\imath} \; \qeq \psi^{\bar\imath} \= 0 \, ,
\end{equation}
which is satisfied based on the fact that the background isometry
$\delta_{\mathring{\xi}}$ is compact.  The deformed action
corresponding to \eqref{eq:quadr-deformation} now takes the form,
\begin{align}
  \label{eq:deformed-action}
  S(\lambda) \=\; & S^\mathrm{class}[\mathring{\phi} + \widetilde{\phi}] 
                    + \int\mathrm{d}^n x \, 
                    \big[ B_\alpha F(\mathring{\phi},\widetilde{\phi})^\alpha 
                    + (-)^{\epsilon_\alpha} b_\alpha\,\qeq F(\mathring{\phi},\widetilde{\phi})^\alpha\,\big] \\
                  &\,+ \lambda \int\mathrm{d}^n x \sum_{\bar\imath}
                    \sqrt{\mathring{g}} \;\big[ \delta_\mathrm{eq}\overline{\psi}_{\bar\imath}  \;
                    \delta_\mathrm{eq}\psi^{\bar\imath}  -
                    \overline{\psi}_{\bar\imath}\;\delta_\mathrm{eq}{\!}^2
                    \psi^{\bar\imath}\,\big]\, \, . \nonumber
\end{align}
Integrating over the Lagrange multiplier fields~$B_\alpha$ in the 
functional integral yields a functional delta function imposing 
the gauge conditions~$F(\mathring{\phi},\widetilde{\phi})^\alpha = 0$.
We can therefore write
\begin{align}
\label{eq:Z-gf}
  Z[\mathring{\phi}] \=& \lim_{\lambda \rightarrow \infty}
                         Z[\mathring{\phi};\lambda] \nonumber \\ 
  \=& \lim_{\lambda \rightarrow \infty} \, \int
      \,\mathcal{D}\widetilde{\phi}^{\,i}\, \mathcal{D}c^\alpha \,
      \mathcal{D}b_\alpha\; \delta\big[F(\mathring{\phi},\widetilde{\phi})^\alpha\big] \,  
      \exp\big[ S^\mathrm{class}[\mathring{\phi} + \widetilde{\phi}] + (-)^{\epsilon_\alpha} b_\alpha\,\qeq 
      F(\mathring{\phi},\widetilde{\phi})^\alpha\,\big] \times \nonumber \\
       &\qquad \qquad \exp\Big[ \lambda \int\mathrm{d}^n x \sum_{\bar\imath}
                  \sqrt{\mathring{g}} \;\big[ \delta_\mathrm{eq}\overline{\psi}_{\bar\imath}  \;
                   \delta_\mathrm{eq}\psi^{\bar\imath}  -
                   \overline{\psi}_{\bar\imath}\;\delta_\mathrm{eq}{\!}^2
                   \psi^{\bar\imath}\,\big]\Big] \, .
\end{align}
Note that in this form, the number of bosonic and fermionic degrees of 
freedom no longer match; we will restore the balance at a later stage.

In the limit $\lambda\rightarrow\infty$ the critical points of the
deformation dominate the functional integral~\eqref{eq:Z-gf}. We
assume that this critical locus is bosonic, i.e.~we can set all the
anti-commuting fields and ghosts to zero. The resulting
\emph{localization manifold} is
\begin{equation} 
  \label{eq:fervar} 
  \mathcal{M} \= \big\{ \qeq \psi^{\bar\imath} = 0
  \;\; \text{for all fermions} \;\psi^{\bar\imath} \in  \widetilde{\phi}^{\,i}\,
  \big{/} F(\mathring{\phi},\widetilde{\phi})^\alpha = 0 \, \big\} 
  \, \equiv \, \big\{t_a \big\} \, , 
\end{equation}
where the parameters~$t_a$ are appropriately chosen 
coordinates on the solution set~$\mathcal{M}$. The localization
manifold includes the gauge-fixing conditions due to the delta
functional in~\eqref{eq:Z-gf}, and~\eqref{eq:fervar} instructs us to
impose the vanishing of the~$\qeq$-variations for \emph{all} the
fermion fields~$\psi^{\bar\imath}$. Therefore the
term~$b_\alpha\,\qeq F(\mathring{\phi},\widetilde{\phi})^\alpha$ in
the undeformed action plays no role in the characterization
of~$\mathcal{M}$.

To appreciate what the consequences are of the conditions
$\delta_\mathrm{eq} \psi^{\bar\imath} =0$, we remind the reader that
the purely bosonic terms of $\delta_\mathrm{eq} \psi^{\bar\imath}$
take the form
$R(\mathring{\phi}+\widetilde{\phi})^{\bar\imath}{\!}_{\bar\alpha}\,\Lambda
(\mathring{c}+c)^{\bar \alpha}$, where the index $\bar{\alpha}$ refers
to fermionic gauge parameters so that their corresponding ghosts are
commuting fields.\footnote{
  We remind the reader that $\Lambda$ is only present to keep track of
  the relative signs between the contributions from fermionic and
  bosonic fields. When writing the various expressions explicitly in
  terms of fermionic and bosonic fields, the presence of $\Lambda$ can
  be avoided.  } 
For a bosonic localization manifold the dependence of
$R(\mathring{\phi}+\widetilde{\phi})^{\bar\imath}{\!}_{\bar\alpha}$ on
the fermion fields is suppressed so that this manifold will involve
the bosonic fields $\widetilde{\phi}^{\,\hat\imath}$, subject to gauge
conditions, and $c^{\,\bar \alpha}$. Both types of fields must vanish
on the boundary. The background fields $\mathring{\phi}^{\,i}$ and
$\mathring{c}^{\,\alpha}$ are subject to the invariance condition
\eqref{eq:isometry}. The background ghosts~$\mathring{c}^{\,\alpha}$ 
that parametrize the background isometries must be restricted such that the square 
of the corresponding~$\qeq$ variation yields a compact~$\delta_{\mathring{\xi}}$ 
(cf. the discussion below~\eqref{eq:Z-eq-inv}). 
The solution of the equations
$R(\mathring{\phi}+\widetilde{\phi})^{\bar\imath}{\!}_{\bar\alpha}\,\Lambda
(\mathring{c}+c)^{\bar \alpha}= 0$ then impose relations between the
fields $\widetilde{\phi}^{\,\hat\imath}$ and $c^{\,\bar \alpha}$ that
lead to the localization manifold. This manifold will be parametrized
in terms of the independent coordinates $t_a$ that we have introduced
in \eqref{eq:fervar}. Not surprisingly, the same type of equations are
encountered when determining supersymmetric field configurations in
classical supergravity, where the ghost fields are replaced by the
parameters of the supersymmetry transformations. There are various
ways to solve such equations, and we will discuss a specific
application in the next section by way of an illustration.

The localization manifold~$\mathcal{M}$ thus corresponds to the set 
of critical points with certain values
for the bosonic fields $\widetilde{\phi}^{\,\hat\imath}$ and
$c^{\bar{\alpha}}$, which we denote by
$\widetilde\phi^{\,\hat\imath}(t)\vert_\mathcal{M}$ and
$c^{\bar{\alpha}}(t)\vert_\mathcal{M}$.
We can then expand the quantum fields as follows
\begin{equation} 
  \label{eq:fluctuations}
  \widetilde{\phi}^{\,i} \= \widetilde{\phi}^{\,i}(t)|_{\mathcal{M}} +
  \tfrac{1}{\sqrt{\lambda}}\,\widetilde{\phi}^{\,i}{\,}' \, , \quad\;
  c^\alpha \= c^{\,\alpha}(t)|_{\mathcal{M}} +
  \tfrac{1}{\sqrt{\lambda}}\,c^\alpha{\,}'\, ,
\end{equation}
where the fermionic fields
$\widetilde\phi^{\,\bar\imath}(t)\vert_\mathcal{M}$ and
$c^{\hat{\alpha}}(t)\vert_\mathcal{M}$ vanish.  As alluded to above,
the anti-ghost fields do not appear in the~$\qeq$-variation of the
fermionic fields~$\psi^{\bar\imath}$ and are therefore not part of the
localization manifold and should be regarded as quantum fluctuations.
To ensure that all propagators scale uniformly with~$\lambda$ in the
expansion~\eqref{eq:fluctuations}, we also rescale the anti-ghosts as
\begin{equation}
  \label{eq:antighostrescale}
  b_\alpha \= \sqrt{\lambda}\,b_\alpha{}' \, .
\end{equation}

With these definitions one can expand the exponent of the integrand in
\eqref{eq:Z-gf} according to \eqref{eq:fluctuations} and
\eqref{eq:antighostrescale}, taking into account that the localization
manifold is purely bosonic. The result is then equal to the classical
action evaluated at the localization manifold and all the terms from
the deformation and the gauge-fixing terms proportional to the
anti-ghosts that are quadratic in the fluctuations
$\widetilde{\phi}^{i}{\,}'$, $c^{\,\alpha}{\,}'$ and $b_\alpha{}'$, up
to terms that vanish in the large-$\lambda$ limit.  Integrating over
these fluctuations then gives rise to the following result for the
functional integral~\eqref{eq:Z-gf},
\begin{equation} 
  \label{eq:Zfinal} 
  Z[\mathring{\phi}] \= \int_\mathcal{M} \, \mu(t)\, \mathrm{d} t_a 
  \, \exp\big[ S^\mathrm{class}[\mathring{\phi},
  \mathring{c}\,;t_a]\, \big] \; Z_\text{1-loop}[
  \mathring{\phi}, \mathring{c}\,;t_a] \, , 
\end{equation}  
where we have assumed the presence of a measure $\mu(t)$ induced by the
embedding of the localization manifold into the field configuration
space. This measure can in principle be evaluated from the explicit
expressions for $\widetilde\phi^{\,\hat\imath}(t)\vert_\mathcal{M}$
and $c^{\bar{\alpha}}(t)\vert_\mathcal{M}$.

The last term $Z_\text{1-loop}[\mathring{\phi}, \mathring{c}\,;t_a]$
under the integral contains the semiclassical correction caused by the
integration over the quantum fluctuations of the fields about the
localization manifold.  These corrections follow from expanding the
exponent in~\eqref{eq:Z-gf} according to
\eqref{eq:fluctuations}, retaining only the terms quadratic in the
fluctuations. In order to keep the balance between fermionic and
bosonic fluctuations manifest, we rewrite the delta functional over the
gauge-fixing terms by reinstating the Lagrange multiplier fields
$B_\alpha{\!}'$, such that~$\qeq b_\alpha{}' = \Lambda
B_\alpha{}'$. In this way we obtain
\begin{align}
\label{eq:Z1loop}
 Z_\text{1-loop}[\mathring{\phi}, \mathring{c}\,;t_a] \= 
 \int\mathcal{D}(\widetilde{\phi}^{\,i}{\,}')&\,\mathcal{D}(c^\alpha{\,}')\,
 \mathcal{D}(b_\alpha{}')\,\mathcal{D}(B_\alpha{}') \, \times \nonumber \\
 &\;\exp\Big[\qeq\big[\mathcal{V} 
 + b_\alpha{}' F(\mathring{\phi}\,;t_a,\widetilde{\phi}{\,}')^\alpha\,\big]\Big]\Big|_{\mathrm{quad.}} \, .
\end{align}
The only contribution to the integrand above comes from 
terms quadratic in the fluctuations, so the gaussian integration over these oscillations 
will lead to a super-determinant. Because the localization manifold is purely
bosonic, this super-determinant is simply equal to the ratio of
two determinants, one associated with the fermionic fluctuations and
the other with the bosonic fluctuations. These determinants can then
be computed by explicit diagonalization, or by making use of powerful
fixed-point formulas~\cite{AtiyahBook:1974}.  Of course, obtaining
explicit expressions must be done in the context of a specific
application.

Let us close this analysis with the remark that 
we have presented the formula~\eqref{eq:Zfinal} including only
the contributions from smooth field configurations.  In addition, one
must also allow for field configurations that are singular precisely
at the fixed point in space-time of~$\d_{\mathring{\xi}}$, which in
super-Yang-Mills theories, for instance, correspond to point-like
instantons~\cite{Nekrasov:2002qd,Pestun:2007rz}.

\section{Application to exact quantum entropy of 
supersymmetric black holes}
\label{sec:application}
\setcounter{equation}{0}

In the previous sections we have been very general about the nature of
the theory that we may wish to consider.  In this closing section we
therefore turn to a specific direction of interest that demonstrates
how our construction of the equivariant algebra naturally lends itself
to computing supersymmetric gravitational functional integrals in
asymptotically $\mathrm{AdS}$ spaces, where the boundary conditions on
the fields are dictated by the conformal boundary of the
space~\cite{Witten:1998qj}. The background can be chosen to be 
supersymmetric  with an $\mathrm{AdS}_n \times S^m$
geometry and the supersymmetry background ghosts are associated with a
particular supercharge on this background characterized by a
generalized Killing spinor. The observables in our BRST cohomology in
this case would be the holographic analogs of protected calculations
in the boundary gauge theory,\footnote{
  Some classical aspects of a special class of such observables have
  been recently discussed in~\cite{BenettiGenolini:2017zmu}; related
  ideas in a slightly different context of topological strings are
  discussed in~\cite{Brennan:2017rbf}.} 
which leads to an exciting possibility for an exact AdS/CFT
correspondence.

To illustrate this idea in a concrete example, we revisit the analysis
of~\cite{Dabholkar:2010uh,Dabholkar:2011ec} of the quantum entropy of
dyonic four-dimensional half-BPS black holes in~$\CN=2$
supergravity in the context of the formalism of this paper. Our construction of
the equivariant algebra~\eqref{eq:equivariant} provides a proper
framework for applying localization of the path integral for
supergravity theories defined on spaces with an asymptotic boundary,
as outlined in Section \ref{sec:localization}. Hence it can in
principle be applied to the path integral that defines the quantum
entropy~\cite{Sen:2008vm}. Further details of actual computations will
appear in a forthcoming paper~\cite{JeonMurthy}, but here our aim is
to present an overview of this application in order to further clarify the
formal discussions of the previous sections.

Let us start by specifying the $\qeq$-variations as derived in
Section~\ref{sec:equiv-cohom} of the most relevant fermion fields in a
purely bosonic field configuration in the context of the superconformal formulation
of $\mathcal{N}=2$ supergravity ~\cite{deWit:1984rvr,deWit:1984wbb}.
These fermions belong to the Weyl and the vector supermultiplets.  Here
we will make use of the off-shell gauge algebra of Euclidean~$\CN=2$
superconformal gravity presented in~\cite{deWit:2017cle}. The Weyl
supermultiplet contains the gravitino fields whose $\qeq$-variation in
a bosonic field configuration equals
\begin{equation}
  \label{eq:qeq-gravitino} 
  \qeq \, \psi_{\mu\,\pm}^i \= 
  2\,\mathcal{D}_\mu(\varepsilon^{i} + {c}_Q{\!}^{i} )_\pm  
  + \tfrac{1}{16}\mathrm{i}\,T_{ab}\,\gamma^{ab}\gamma_\mu
  (\varepsilon^{i} + {c}_Q{\!}^{i} )_\mp  
  - \mathrm{i}\,\gamma_\mu\,(\eta^i + {c}_{S}{\!}^{i} )_{\mp} \; , 
\end{equation}
where the subscript~$\pm$ on the fermions and ghosts denote chiral
projections. The quantum ghosts associated with $Q$- and
$S$-supersymetry are $c_Q{\!}^{i}$ and~$c_S{}^{i}$ and the
corresponding background ghosts are $\varepsilon^i$ and
$\eta^i$. Here and henceforth the indices $i, j, \dots$ refer
to the $\mathrm{SU}(2)$ R-symmetry. Note that we have suppressed the
universal anticommuting parameter $\Lambda$ since, in a bosonic field
configuration, there are no subtleties with relative signs of the various contributions. The other bosonic fields in these equations
are the metric, the auxiliary tensor~$T_{ab}$ as well as related gauge
connections that are part of the off-shell Weyl multiplet. All these
fields must be decomposed into background and quantum fields,
as we have explained in previous sections. The off-shell Weyl
multiplet also contains another fermion field, which we will ignore
here because it only plays a minor role in what follows.

To describe the electric and magnetic charges of the black hole, the
supergravity must include a number of vector supermultiplets labeled
by~$I=0\ldots n_\mathrm{v}$. Their corresponding
fermions~$\Omega^{I\,i}$ have the following $\qeq$-variation in a
bosonic field configuration
\begin{equation}
   \label{eq:qeq-gaugino}
  \qeq\,\Omega^{I\,i}_\pm \=  2\mathrm{i}\,\Slash{\mathcal{D}}
                               X_\pm^I 
     \,(\varepsilon^{i} + {c}_Q{\!}^{i} )_\mp 
   + \big[\tfrac12\,\delta_j{}^i \widehat{F}_{ab}^{I\,\mp}\gamma^{ab}
  + \varepsilon_{kj}Y^{I\,ik}\big](\varepsilon^{j} + {c}_Q{\!}^{j} )_\pm 
    - 2\,X_\pm^I\,(\eta^i + {c}_{S}{\!}^{i})_\pm    \; .
\end{equation}
The right-hand side of the above equation contains the real scalar
fields~$X_\pm{\!}^I$ and the auxiliary~$\mathrm{SU}(2)$
triplets~$Y^{I\,ij}$, whereas the gauge fields enter through the
(anti-)selfdual projections of the modified field
strength~$\widehat{F}^I_{ab}$.  The covariant derivative on the
scalars contains the various connections belonging to the Weyl
multiplet.

Although the equations~\eqref{eq:qeq-gravitino} and
\eqref{eq:qeq-gaugino} represent the {\it equivariant} variations
$\qeq$ of the quantum fermions, they reduce to the $Q$- and
$S$-supersymmetry transformations of the fermions prior to the
background field split upon identifying the fields on the right-hand
side with the background fields and, at the same time, suppressing the
quantum ghosts $c_Q{\!}^i$ and $c_S{}^i$, and keeping only the
background ghosts $\varepsilon^i$ and $\eta^i$. Hence they can be used
to exhibit the consequences of full supersymmetry for the near-horizon
geometry. Since the background must be fully supersymmetric the
truncated equations~\eqref{eq:qeq-gravitino} and
\eqref{eq:qeq-gaugino} must vanish for all values of the background
ghosts $\varepsilon^i$ and $\eta^i$ (up to certain gauge choices). The
result of this analysis is that the background geometry must be
$\mathrm{AdS}_{2} \times S^{2}$ and that the full background is
invariant under eight supersymmetries generated by particular linear
combinations of $\varepsilon^i$ and $\eta^i$ that define eight
independent Killing spinors associated with the fermionic isometries
of the full background configuration. Such spinors are not
normalizable in the asymptotic $\mathrm{AdS}_2$ space.

Obviously the above analysis leads precisely to the fully
supersymmetric near-horizon geometry $\mathrm{AdS}_{2} \times S^{2}$
with fixed electric and magnetic
fluxes~\cite{LopesCardoso:1998tkj,LopesCardoso:2000qm}.  Here we
should add that the background values of the gauge fields are
constrained by the background values of the scalars $X_\pm^I$.  The
gauge fields carry fixed electric and magnetic charges, corresponding
to the microcanonical ensemble. The condition of fixed magnetic
charges is implemented on the gauge field components along the~$S^2$
in the asymptotic region. The condition of fixed electric charges is
implemented in the classical theory by a Legendre transform with
respect to the electric fields.  In the quantum theory this requires
the introduction of a Wilson line at the boundary of the near-horizon
region, and we must compute the expectation value of this operator by
integrating all fluctuations of all the supergravity fields around the
above background~\cite{Sen:2008vm}.

Now we turn to the computation of the functional integral, by
following the localization procedure explained in
Section~\ref{sec:localization}.\footnote{
  For $\mathrm{AdS}_2$ one has to take into account an additional
  subtlety coming from the fact that there are normalizable gauge
  transformations with corresponding non-normalizable gauge parameters 
  \cite{Banerjee:2009af}.} 
For that purpose we have to determine the localization manifold, which
follows from requiring the $\qeq$-variations of the quantum fermions,
given by \eqref{eq:qeq-gravitino} and \eqref{eq:qeq-gaugino}, to
vanish.  Recall from the comment
following~\eqref{eq:truncate-brst-quantum} that our formalism still
allows us to restrict the number of background ghosts parametrizing
the~$\qeq$-variations. Thus, we can reduce the final problem to
finding all geometries and bosonic matter field configurations that
asymptote to the near-horizon background, and admits a Killing spinor
which asymptotes to a particular near-horizon Killing spinor. This
particular Killing spinor can be chosen so that the square of the
equivariant variations generates a background transformation
$\delta_{\mathring{\xi}} = L - J$, where $L$ and $J$ are compact
$\mathrm{U}(1)$ rotations of the $\mathrm{AdS}_2$ and $S^2$ factors in
the background geometry, respectively.  This is precisely the problem
addressed in~\cite{Dabholkar:2010uh} and solved in~\cite{Gupta:2012cy}
for smooth field configurations. In order to complete the calculation
of the functional integral~\eqref{eq:Zfinal}, we then need to evaluate
the physical action of the theory on these localizing
configurations~\cite{Dabholkar:2010uh,Dabholkar:2011ec,Murthy:2013xpa},
and we need to compute the one-loop fluctuation 
determinant~\cite{Murthy:2015yfa, Gupta:2015gga, JeonMurthy}.
Finally, as mentioned above, the smoothness assumption that we made in
supergravity should be removed in string theory, wherein a class of
orbifold configurations also contribute to the functional
integral~\cite{Banerjee:2008ky,Murthy:2009dq}.

\vspace{0.4cm}

It is clear that the quantum entropy problem for BPS black holes in
asymptotically flat space is but one application of our ideas. The
formalism constructed in this paper~is~quite~general~in~that~it~can~be
defined~around~an~arbitrary~background~that~admits~(super-)isometries.
Our discussion gives a precise physical realization of the idea of
equivariant cohomology, and of the corresponding equivariant
localization using the background supersymmetry ghosts, in the
variables of supergravity. We hope that the framework 
outlined in this paper will prove useful
in a variety of other physical situations.

\section*{Acknowledgements}
We thank Atish Dabholkar, Camillo Imbimbo, Rajesh Gupta, Imtak Jeon,
Paul Richmond and Alberto Zaffaroni for interesting and useful
discussions. B.~de Wit and V.~Reys thank King's College London for
hospitality during the course of this work. The work of S.~Murthy was
supported by the ERC Consolidator Grant N.~681908, ``Quantum black
holes: A macroscopic window into the microstructure of gravity'', and
by the STFC grant ST/P000258/1.  V.~Reys is supported in part by INFN
and by the ERC Starting Grant 637844-HBQFTNCER.


\end{document}